\documentclass[unpublished, a4paper]{quantumarticle}
\pdfoutput=1
\usepackage{graphicx}
\usepackage[utf8]{inputenc}
\usepackage{amsmath}
\usepackage{float}
\usepackage{booktabs}
\usepackage[ruled]{algorithm2e}
\usepackage[noend]{algpseudocode}

\renewcommand{\thefootnote}{\arabic{footnote}}

\title{Boosting the Performance of Quantum Annealers using Machine Learning}

\author{Jure Brence*\textsuperscript{1,2,6}, Dragan Mihailović\textsuperscript{3,4,5}, Viktor V. Kabanov\textsuperscript{3},\\ Ljupčo Todorovski\textsuperscript{1,6}, Sašo Džeroski\textsuperscript{1,2}, Jaka Vodeb\textsuperscript{3,5}}
\date{March 4, 2022}

\begin{document}

\maketitle
\let\thefootnote\relax\footnote{* Corresponding author, email: jure.brence@ijs.si\\
\textsuperscript{1} Department for Knowledge Technologies, Jozef Stefan Institute, Jamova 39, 1000 Ljubljana, Slovenia\\
\textsuperscript{2} Jozef Stefan International Postgraduate School, Jamova 39, 1000 Ljubljana, Slovenia\\
\textsuperscript{3} Department of Complex Matter, Jo{\v z}ef Stefan Institute, Jamova 39, 1000 Ljubljana, Slovenia\\
\textsuperscript{4} CENN Nanocenter, Jamova 39, 1000 Ljubljana, Slovenia\\
\textsuperscript{5} Department of Physics, Faculty for Mathematics and Physics, Jadranska 19, University of Ljubljana, 1000 Ljubljana, Slovenia\\
\textsuperscript{6} Department of Mathematics, Faculty for Mathematics and Physics, Jadranska 19, University of Ljubljana, 1000 Ljubljana, Slovenia
}

\begin{abstract}
Noisy intermediate-scale quantum (NISQ) devices are spearheading the second quantum revolution. Of these, quantum annealers are the only ones currently offering real world, commercial applications on as many as 5000 qubits.
The size of problems that can be solved by quantum annealers is limited mainly by errors caused by environmental noise and intrinsic imperfections of the processor.
We address the issue of intrinsic imperfections with a novel error correction approach, based on machine learning methods. Our approach adjusts the input Hamiltonian to maximize the probability of finding the solution.
In our experiments, the proposed error correction method improved the performance of annealing by up to three orders of magnitude and enabled the solving of a previously intractable, maximally complex problem.
\end{abstract}

\maketitle

\section{Introduction}
Quantum annealers have proven themselves useful in solving various problems in material science \cite{Harris2018,King2018,Bando2020,Vodeb2021,Kairys2020}, optimization \cite{Neukart2017,Orus2019} and  machine learning \cite{Mott2017,Li2018,Willsch2020,Jain2020}, and have shown scaling advantage in problem solving efficiency in some cases \cite{King2021,Albash2018}. 
However, the quantum computers used in these cases have trouble excluding the impact of device imperfections and the outside environment on the quantum dynamics taking place within the quantum device \cite{Job2018,Bando2020,Boixo2016,Gardas20181,Gardas20182}. The resulting errors limit the potential of quantum simulations or quantum speed-up in solving classical optimization problems.
Quantum error correction assumes two very different realizations in the two quantum computing models currently used in practice. The quantum gate model (e.g., Google and IBM) already has an established quantum threshold theorem, which states that an arbitrarily long quantum computation circuit can be constructed, provided the qubits involved have a low enough error rate. However, quantum annealing has no such established theorem. Error correction algorithms are therefore scarce and so are examples of research on quantum annealing error correction and topology compensation. 
The problem of noise in quantum annealing has so far been tackled only in two different ways that do not involve improvements to hardware. The first is quantum annealing error correction, based on introducing an energy penalty along with encoding and error correction \cite{Pudenz2014}. The second approach involves compensating for differences in chain susceptibility due to the embedding topology \cite{Raymond2020}.

In this paper, we address the issue of noise in D-Wave's 2000Q quantum annealer. The way a problem is solved on this device is to package the problem at hand into a Hamiltonian, employing the quadratic unconstrained binary optimization (QUBO) formalism
\begin{equation}
    \label{eq1}
    H=\sum_{i,j}Q_{i,j}q_iq_j,
\end{equation}
where specifying the matrix $Q$ corresponds to packaging up the problem as input to the quantum processing unit (QPU). The elements $Q_{i,j}$ are couplers if $i\neq j$ and biases if $i=j$. $q_i$ represents a logical qubit (LQ) with the possible values of $0$ and $1$, which is used in order to interpret the output solution provided by the QPU. There is an important distinction between LQs and physical qubits (PQs). PQs are non-abstract actual physical qubits, which are manipulated by the couplings and biases applied to them during the quantum annealing process in the form of physical magnetic fields on the chip. Their connectivity is fixed into a so-called Chimera graph \cite{TechnicalQPU}, where each PQ is connected to $6$ neighbors. However, the connectivity which is required between LQs by the matrix $Q$ is rarely the same as the Chimera graph. In order to realize the desired connectivity, additional PQs and couplings are added, so that each LQ is represented by a chain, composed of several PQs. This mapping between LQs and PQs is called an embedding \cite{TechnicalQPU}. 

\begin{figure*}
        \centering
        \includegraphics[width = 0.9\textwidth]{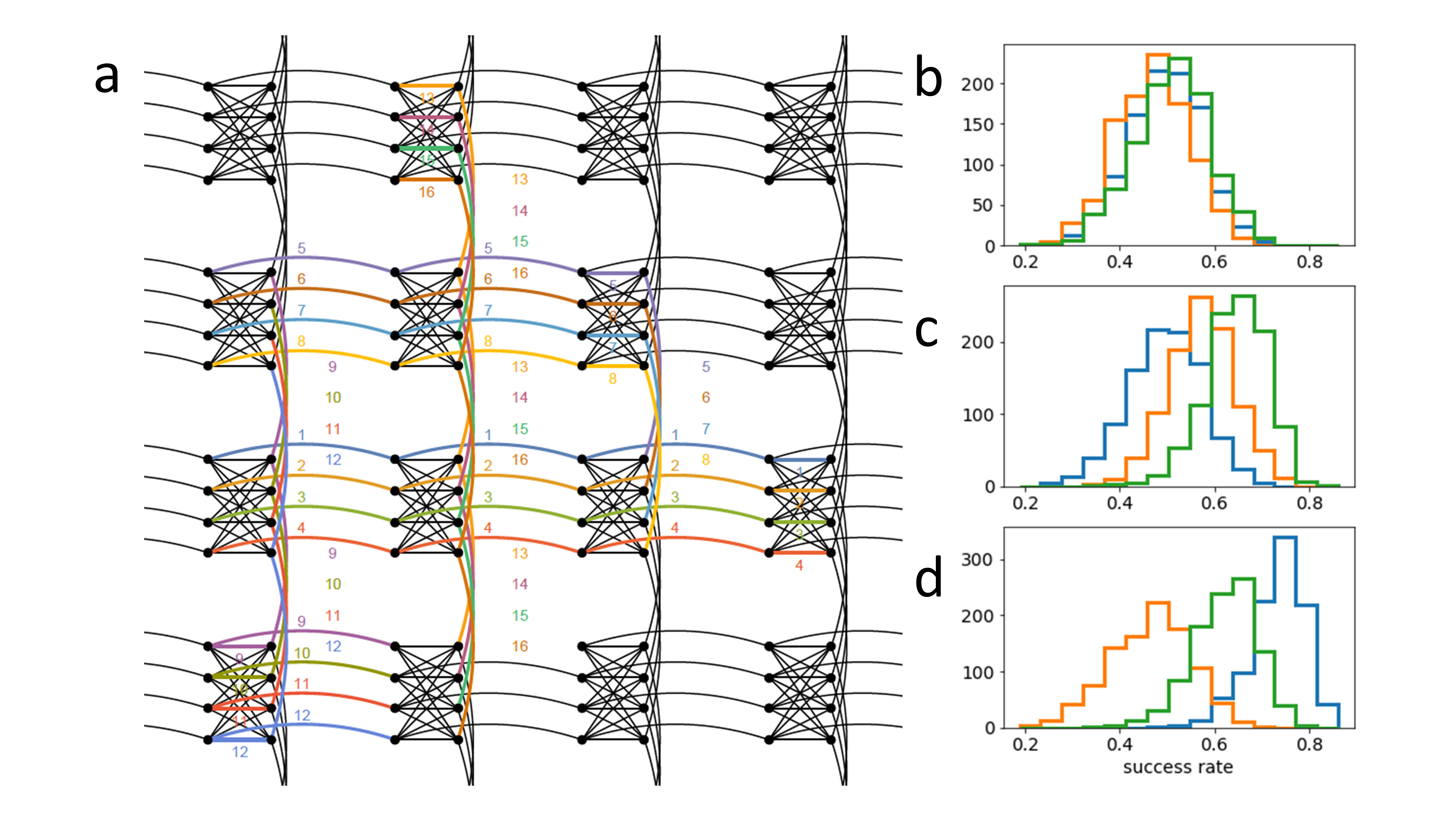}
        \caption{\textbf{a)} An example of an embedding. Physical qubits are represented with black dots and the connections between them with black lines. Logical qubits are enumerated and shown with colored connections between physical qubits. Each of the colored connections is forcing the two physical qubits to assume the same value. Therefore, a logical qubit is a chain of physical qubits, all (or most) of which assume the same value. There are post-processing methods in place on the quantum computer, which take care of disagreements between physical qubits in a chain. \textbf{b)} Distributions of the success rate for three samplings of size 1000, when using the same embedding for the three samplings, \textbf{c)} when using translated embeddings, and \textbf{d)} when using distinct embeddings.}
        \label{fig1}
\end{figure*}

There are many different sources of noise present on the QPU, as specified in the documentation provided by D-Wave: $(i)$ background susceptibility which comes from next-nearest neighbor interactions and leakage of biases, $(ii)$ $1/f$ flux noise is exerted on qubits which manifests itself as a drift of their properties on a larger time-scale, such as between different problem submissions, $(iii)$ the problem Hamiltonian, which is specified as part of the input to the machine has a finite resolution, $(iv)$ the ratio between biases and couplings can vary for different annealing parameters, and $(v)$ qubits cannot be made perfectly identical. In addition, several studies have revealed that the QPU is inherently coupled to its environment, which brings in $(vi)$ thermal effects \cite{Albash2015,Benedetti2016,Boixo2016,Buffoni2020} and $(vii)$ random quantum fluctuations in the Josephson current present in qubits \cite{Bando2020}. We take all of these effects into account by modelling the noise on the QPU as
\begin{equation}
    \label{eq2}
    Q_{\text{submitted}} = Q_0 + dQ_{\text{calibration}} + dQ_{\text{noise}},
\end{equation}
where each element of $dQ_{\text{noise}}$ is a Gaussian random variable with mean $0$ and standard deviation $\sigma$. Elements of $dQ_{\text{calibration}}$ represent the shifts of the mean of the Gaussian random variables. The idea of this work is to find the appropriate $dQ_{\text{calibration}}$, which we then subtract from $Q_{\text{submitted}}$. We assume that $dQ_{\text{calibration}}$ is constant and represents the combined systematic noise from all the aforementioned noise sources. The degree to which we can reduce the impact of noise is presently unknown and determining it the motivation of this investigation.

\section{Results}
In order to study the effectiveness of our error correction algorithms, we define two metrics based on the energies of the output samples from the QPU. A single output sample is a string of $0$s and $1$s representing the final values of $q_i$. The energy $H$ can then be calculated via Eq. \ref{eq1}. The first metric is the success rate, defined as the ratio between the number of samples with the lowest possible energy (ground state) and the total number of samples in a single submitted problem. The second metric is the mean energy, which is simply the mean of all the energies in a sample set. We use these two metrics interchangeably, depending on the specific problem, because either maximizing the success rate or minimizing the mean energy achieves the same goal of improving annealing performance. See SI, section~1 for details.

The physical problem of finding the ground state of a system of electrons on a triangular lattice was studied previously without error correction \cite{Vodeb2021}. The studied physical problem requires maximal connectivity between LQs, which is represented by a fully connected graph or a clique. Such a connectivity, when embedded onto D-Wave's Chimera graph of PQs, is one of the most complex problems one can attempt to solve on a quantum annealer. Therefore, if we are able to show that our error correction algorithms work in this case, then they are robust and will work well also with other complex problems. 

In the D-Wave documentation, it is recommended that users find the best embedding for the specific problem that they want to solve by trial-and-error. In Fig.~1a, we show an example of a fully connected graph embedded into a Chimera graph. Figs.~1b-d demonstrate the effect of changing the embedding. Fig.~1b shows what happens when we simply use the same embedding in three different samplings, each consisting of 1000 problem submissions to the QPU. The three resulting distributions of success rates are consistent over time and demonstrate reproducibility. Fig.~1c shows the results of sampling when the embedding is translated across the physical qubits for the three different samplings (Fig.~1c), while Fig.~1d shows the results when the embedding is entirely changed for the three different samplings. The two examples show that both the specific topology and the position of the embedding influence the success rate. Therefore, our error correction algorithm needs to address each individual coupler and bias in $Q_0$. For details of our study of annealing offsets see SI, section~2 and Supplementary Figure~1.

In order to address each coupler and bias individually, we define our correction $dQ$ to the problem Hamiltonian $Q_0$ and submit to the QPU the sum of the two as $Q_{\text{submitted}}=Q_0+dQ$. We assume that, given a large enough data set, $dQ\rightarrow -Q_{\text{calibration}}$, which effectively cancels out the systematic noise present on the device. However, due to practical limitations of a finite size data set, we constrained the scope of our search of $dQ$. When determining each $dQ_{i,j}$, we define the upper and lower bound to this parameter so that the corrected element $Q_{\text{submitted}i,j}=Q_{0i,j}+dQ_{i,j}$ deviates only by a maximum fraction $\eta$ from the original $Q_{0i,j}$. The value of $\eta$ should be between $0\leq\eta\leq1$. It is important that the value is kept small, lest we risk changing the original physical problem into an entirely different one. We choose a value of $\eta=0.05$ unless specified otherwise. For more details see SI, sections 3 and 4.

We study three sizes of the triangular lattice on which the electrons reside ($4x4$, $5x5$ and $6x6$ atomic sites). These lattice sizes correspond to $136$, $325$ and $666$ free parameters to determine in $dQ$, respectively. We expect that as the number of parameters increases, the complexity of finding appropriate corrections also increases.

We evaluate two adaptive quantum annealing correction algorithms. The first and simpler of the two is called the \textit{argmax} strategy, summarized in Algorithm~\ref{alg:argmax}.

\begin{algorithm}
\caption{The \textit{argmax} strategy of adaptive quantum annealing correction}
\begin{algorithmic}[1]
  \State Compute the ranges for each $dQ_{ij}$ for the problem $Q_0$ and choose $\eta$.
  \State Sample $M$ calibration matrices $dQ$ using a random sampling method (such as latin hypercube sampling \cite{LHS}).
  \State Evaluate the annealing performance for each calibration matrix on the quantum annealer,
  \State Choose the calibration matrix that yields the highest success rate or the lowest mean relative energy.
\end{algorithmic}
\label{alg:argmax}
\end{algorithm}
 To evaluate the \textit{argmax} strategy, we sample $M=10^4$  calibration matrices for each of the three system sizes using latin hypercube sampling. The data is summarized in Supplementary Figure 2.  
 We observe that, for all system sizes, a good fraction of evaluated calibration matrices achieve better annealing performance than the original matrix. This observation is surprising, since we expected it to be very difficult to find a calibration that improves performance in the high dimensional $dQ$ spaces. Following step 4 of the \textit{argmax} strategy, we arrive at the results summarized in Table \ref{tab1}.

\begin{figure*}
        \centering
        \includegraphics[width = 0.75\textwidth]{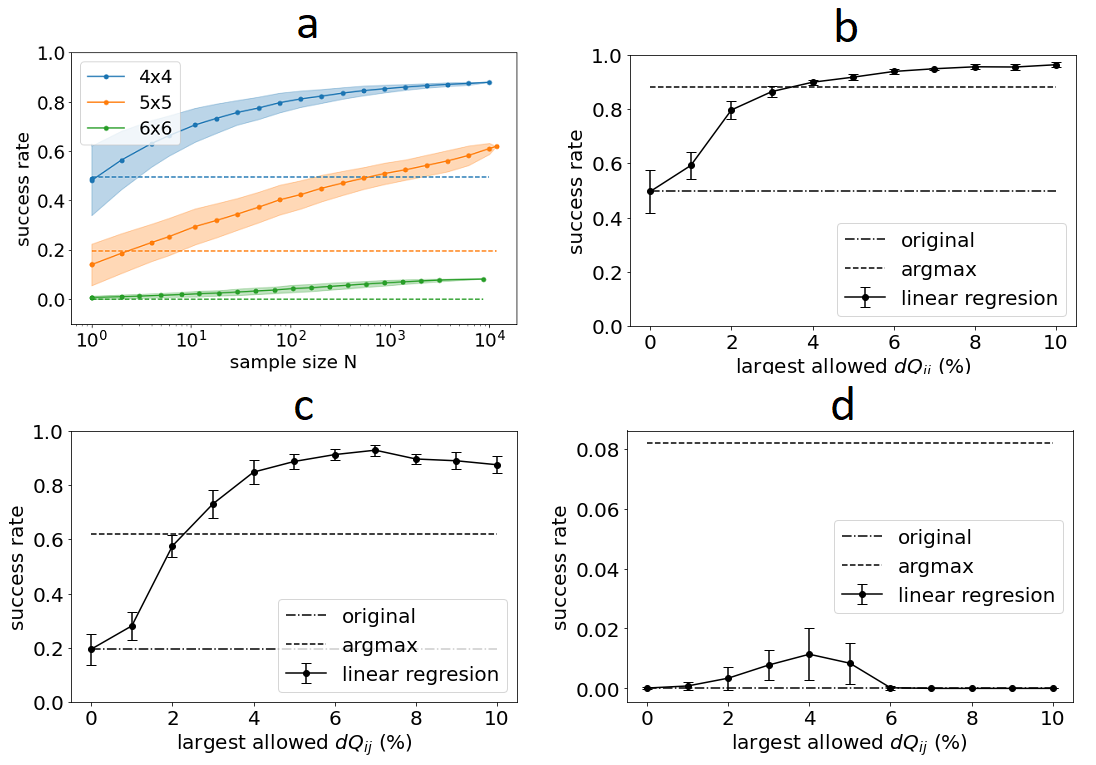}
        \caption{\textbf{a)} Dependence of the success rate of the \textit{argmax} calibration strategy on the size of the training dataset (circles and full line). Calculated through bootstrap resampling with 1000 repetitions. The results are compared with the success rate without calibration (dashed line). Standard deviations of success rate are indicated by the shaded areas. Subfigures b,c,d show the dependence of the success rate of the \textit{linear regression} calibration strategy for the \textbf{b)} 4x4, \textbf{c)} 5x5 and \textbf{d)} 6x6 systems on the largest element of the calibration matrix $dQ_{ij}$ allowed (circles and error bars). For each system size, the results are compared with the success rate of the \textbf{argmax} strategy (dashed line) and with the success rate without calibration (dotted line).}
        \label{fig2}
\end{figure*}

\begin{table}
   \centering
   \caption{Annealing performance, measured by success rate (SR), when using the original $Q_0$, compared to the best annealing performance measured in the sample of 10000 calibrated matrices $Q = Q_0 + dQ$, which corresponds to the \textit{argmax} strategy. \label{tab1}}
   \setlength\tabcolsep{4 pt}
   \begin{tabular}{lrrr}\\
        \toprule
       	  \textbf{system} & \text{4x4} & \text{5x5} & \text{6x6} \\
       	  \midrule
       	  \textbf{SR of original $Q_0$} & 0.50 & 0.20 & 0.0001 \\
       	  \textbf{best SR of $dQ$ sampling} & 0.88 & 0.62 & 0.08 \\
       	  \textbf{\% improvement in SR} & 78\% & 210\% & 80000\% \\
       	\bottomrule
   \end{tabular} 
\end{table}
In order to assess how the strategy behaves with a lower number of samples, we perform repeated bootstrap subsampling on the gathered data (see SI for details) and depict the resulting curves in Figure \ref{fig2}a. We observe that significant improvements in annealing performance can be achieved even for sample sizes as small as $10$ or $100$ calibration matrices. Increases in sample size offer diminishing returns, as the curves asymptotically approach the maximum of the full sample.

The \textit{argmax} strategy is limited to the calibration matrices we have sampled and evaluated. In the following, we employ machine learning methods to develop a strategy that interpolates or even extrapolates the assembled data in order to find the best calibration matrix. 
We train and evaluate models by performing standard 5-fold cross validation (see SI for details) and compare the coefficient of determination $R^2$ of different methods \cite{ESL}.

\begin{table}
   \centering
   \caption{The coefficient of determination $R^2$ achieved by different models for the task of predicting the mean energy, given calibration matrices $dQ$, for three different system sizes. A higher $R^2$ indicates a better model, with the maximal possible value being $R^2 = 1$. The machine learning methods used are linear regression (LR), linear regression with L1 regularization (Lasso), linear regression with L2 regularization (Ridge), support vector machines with an RBF kernel (SVR), ensembles of decision trees (random forest - RF) and feedforward neural networks (FNN).}
   \setlength\tabcolsep{4 pt}
   \begin{tabular}{lrrrrrr}\\
        \toprule
       	  & \textbf{LR} & \textbf{Lasso} & \textbf{Ridge} & \textbf{SVR} & \textbf{RF} & \textbf{FNN} \\ \midrule
	    \textbf{4x4} & $0.475$ & $\textbf{0.476}$ & $0.475$ & $0.454$ & $0.213$ & $0.463$\\
       	\textbf{5x5} &  $0.272$ & $0.274$ & $0.273$ & $0.134$ & $0.067$ & $\textbf{0.277}$\\
       	\textbf{6x6} & $0.091$ & $\textbf{0.113}$ & $0.109$ & $0.076$ & $0.019$ & $0.076$\\
        \bottomrule
   \end{tabular} 
    \label{tab2}
\end{table}
In Table \ref{tab2} we report the $R^2$ for ordinary linear regression -- LR, linear regression with L1 regularization -- Lasso \cite{lasso}, linear regression with L2 regularization -- Ridge \cite{ridge}, support vector regression with a kernel of radial basis functions -- SVR \cite{SVR}, ensembles of decision trees for regression -- random forests -- RF \cite{RF}, as well as feedforward neural networks -- FNN \cite{schmidhuber}. The details of the employed methods are reported in SI.
The coefficients of determination are not impressive, which is consistent with the large variance of the distributions observed in Figure \ref{fig1}. For all system sizes, all variations of linear regression, as well as SVR and FNN, achieve similar predictive performance. The predictive performance drops as we increase system size. This is expected, due to the curse of dimensionality -- the dimensionality of the problem increases quadratically with system size, but our sample size remains constant \cite{ESL}.

\begin{algorithm}
\caption{The \textit{predictive} strategy of adaptive quantum annealer correction.}
\begin{algorithmic}[1]
  \State Compute the ranges for each $dQ_{ij}$ for the problem $Q_0$ and choose $\eta$.
  \State Sample $N$ calibration matrices $dQ$ using a random sampling method (such as latin hypercube \cite{LHS} sampling).
  \State Evaluate the annealing performance for each calibration matrix on the quantum annealer.
  \State Learn predictive models on the data (such as linear regression), employing cross-validation to select the best model.
  \State Vary $\eta$ to define search spaces of increasing size and use an optimization algorithm (such as differential evolution) on the trained model to find the candidate calibration matrix for each $\eta$.
  \State Test the suggested matrices on the quantum annealer and select the best one.
\end{algorithmic}
\label{alg:predictive}
\end{algorithm}

Since linear regression is the simplest and most robust of the models we have tried and still achieved competitive predictive performance, we have chosen it for implementing the \textit{predictive} strategy, summarized in Algorithm~\ref{alg:predictive}. 

We evaluated the \textit{predictive} strategy on the quantum annealer for different values of $\eta$ and depict the results in Figs.~\ref{fig2}b,c,d (see SI for details).
Despite the relatively low coefficients of determination, the proposed strategy shows large improvements in annealing performance. The \textit{predictive} strategy outperforms the \textit{argmax} strategy for the $4x4$ system by 14\% and for the $5x5$ system by 155\%, but falls short of it for the 6x6 system. 
The observations are easy to understand if we consider the fact that for the $4x4$ system, the \textit{argmax} strategy was able to achieve nearly perfect annealing performance, leaving little space for improvement by the \textit{predictive} strategy. For the $6x6$ system, the sample size is too low for the model to make meaningful predictions and it is better to simply use the best evaluated calibration matrix. The $5x5$ system lies in between the other two, where the predictive model learns enough to dramatically outperform the \textit{argmax} strategy. We expect that with more invested QPU time, similar improvements could be achieved for the $6x6$ system as well. See Supplementary~Figure~2 for an estimation of the learning curve of the \textit{predictive} strategy, similar to the performance curve for the \textit{argmax} strategy in Fig.~1a.

Finally, note that the first three steps of the \textit{predictive strategy} are identical to the \textit{argmax} strategy, so the dataset constructed by sampling the $dQ$ space can be reused for both strategies. In fact, we recommend the use of both strategies, because it is difficult to predict which of them will be better for a given problem, as evidenced by our experiments on systems of different sizes. 

\section{Discussion}
\begin{figure}
        \centering
        \includegraphics[width = 0.45\textwidth]{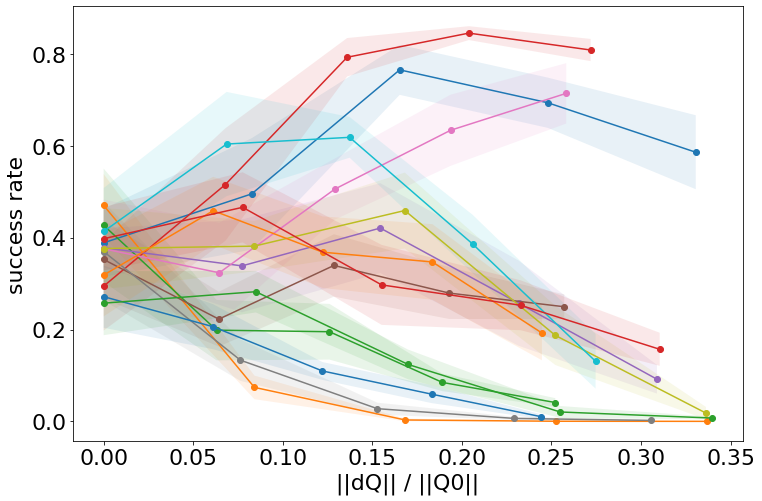}
        \caption{Annealing performance for walks along straight lines in random directions in the $dQ$ space, starting at $Q_0$, for the $4x4$ system. Each color denotes a distinct random direction. Each measurement is repeated 10 times, with the dots and full line depicting the mean and the shaded areas depicting the standard deviation of each experiment.}
        \label{fig3}
\end{figure}
When analyzing the dataset constructed by randomly sampling the $dQ$ space, a few observations stand out, besides the impressive performance of both introduced calibration strategies. Firstly, for all system sizes, nearly half of the sampled calibration matrices result in better annealing performance than using no calibration at all. Secondly, linear regression is able to model the dependence of annealing performance on $dQ$ equally well as more complex, nonlinear methods. The latter can be partially explained by the large amount of random noise present in the measurements, which can increase the risk of overfitting in models, and neccessiate the use of simple, robust models. However, in such a case, one would expect model regularization to improve the predictive performance, which we observed only to a small degree. Therefore, in combination with the former observation, we are lead to the conclusion that the systematic error landscape is much simpler than we had anticipated. 
When considering the success rate, dependent upon the elements of the $dQ$ calibration matrix, a single global peak in the high-dimensional space, with few or no local maxima, would explain our observations. 

To further explore the noise landscape, we perform a new experiment on the $4x4$ system where we perform walks, each starting at $Q_0$ and following a line in randomly chosen direction. The results are depicted in Figure~\ref{fig3}. 
We observe two distinct patterns in the measured walks: the annealing performance either monotonically falls towards zero, or it reaches a single maximum before falling off. 
Furthermore, out of the 14 walks, 4 achieve at some point an annealing performance significantly better than without calibration. 
Were the noise landscape rugged, we would expect the walks to be rugged as well -- experiencing many local extrema before falling towards zero. Furthermore, we would expect walks that improve the starting annealing performance to be extremely rare in a 136-dimensional space. Therefore, the presented observations are consistent with the conclusions we reached when analyzing the $dQ$ dataset. The landscape of systematic error is relatively smooth and likely features only a single maximum for $||dQ|| \ll ||Q_0||$.

This is consistent with a linear approximation of the systematic noise already presented in the D-Wave documentation, where each individual coupling between physical qubits, as well as the external field on each individual qubit, is perturbed by a small constant. Using our calibration methods, we were able to successfully find the appropriate compensation for the specific systematic error present in a specific problem deployed on the D-Wave quantum annealer.

\section{Conclusions}
We have demonstrated a novel approach to quantum annealing error correction, based on compensating for systematic noise by correcting the input to the quantum annealer. Both the \textit{argmax} and the \textit{predictive} strategy achieve significant improvements in annealing performance and can be used as a plug-in module and put to good use by practitioners who struggle with the effects of noise in their quantum annealing experiments and applications. 

The observed properties of the noise landscape hint towards calibration strategies that are more efficient in the use of QPU time.
Gradient-based optimization methods are very efficient, but can easily get stuck in local minima. Since the noise landscape appears to be devoid of local minima, gradient-based methods might prove a good choice. However, since the gradient is not known analytically, it must be estimated at each step of the optimization process by performing measurements on the quantum annealer. In a high-dimensional space, this might prove expensive.
Our observations hint towards the absence of influence of the interactions between the elements of $dQ$ on annealing performance. This property can be exploited by simplifying the problem from optimizing all the dimensions at the same time, to optimizing each of them independently. A strategy of sequential one-dimensional optimization would scale much better with system size and allow for scalable calibration of larger systems.

\section{Data availability}
The data generated and analysed in the presented study are available from the corresponding author on reasonable request.

\section{Acknowledgements}
We acknowledge funding from ARRS projects P2-0103 and P-0040, as well as young researcher grant P08333.

\section{Competing interests}
The authors declare that there are no competing interests.

\section{Authors' contributions}
Jure Brence: investigation, formal analysis, writing -- original draft; Dragan Mihailovic: supervision, funding acquisition, writing -- review and editing; Viktor V. Kabanov: validation, supervision; Ljupčo Todorovski: methodology, validation, writing -- review and editing; Sašo Džeroski: supervision, funding acquisition, writing -- review and editing; Jaka Vodeb: conceptualization, methodology, writing -- original draft.

\clearpage

\section*{Appendix}
\renewcommand{\figurename}{Supplementary Figure}
\setcounter{figure}{0}

\subsection*{Annealing performance}
Because the result of a single quantum annealing experiment is stochastic, annealing is typically repeated hundreds or thousands of times, producing a sample of reads. Since the goal is usually to find the global minimum of the energy, generally only the read with the lowest energy is of interest. We consider the performance of the annealer to be better when a larger fraction of the reads finds the ground state.
In order to summarize the performance of the annealer in a particular experiment from the sample, we look at three quantities:
\begin{itemize}
\item Success rate: the fraction of reads with energy equal to the energy of the ground state.
\item Mean energy: the mean across the energies of the entire sample.
\item Chain break fraction: the fraction of reads with broken chains where not all qubits in a chain end up in the same state.
\end{itemize}
In the process of calibrating the quantum annealer, we seek to maximize the success rate and minimize the mean energy and chain break fraction. The three quantities are highly (inversely) correlated and can be used interchangeably in many cases. 
If the ground state of the system is not known, the success rate can not be calculated. In that case the simplest solution is seeking to minimize the mean energy of the entire sample. However, if we know the ground state of the system, or at least its energy, we can calculate the relative energy of each read, which makes the interpretation of results easier. If the state vector of a given sample is $x$ and the ground state of a Hamiltonian Q0 (in QUBO form) is labelled $x_0$, the energy of the state x can be calculated as

$$E(\textbf{x}) = \textbf{x}^T Q_0 \textbf{x}.$$

We can now define the relative energy of a given quantum state as
$$E_r(\textbf{x})=E(\textbf{x})-E(\textbf{x}_0)$$
and the success rate can be calculated as the fraction of reads with relative energy zero (or close to it, to account for rounding errors).

\subsection*{Embeddings and noise}
In this section, we give detail on the exploratory experiments reported in Figure 1b-d. 
We give insight into the reproducibility of experiments on a quantum annealer. Furthermore, we illustrate the effect of different embeddings on the performance of annealing. All experiments in this section were performed on the 4x4 system under identical settings.

\subsubsection*{Statistics of annealing performance}
Generally, when performing experiments on a quantum annealer, users perform each single experiment with several hundred reads and take the best result. In our work, we are studying the performance of this procedure, quantified through the success rate. To that end, we wish to study the distribution and statistical properties of the success rate, when such an experiment is repeated many times.
We repeat the experiment 1000 times, each time reading 500 quantum states. We calculate the success rate of each experiment and thus obtain 1000 samples of the success rate.
However, due to variable external conditions affecting noise in the quantum system, the performance of annealing may vary when repeated at different times. To account for this, we repeat the sampling of 1000 experiments two more times, with a delay of around 30 minutes between sampling. A histogram for each of the three samplings is depicted in Figure 1b, where different colors correspond to different samplings. The distributions are symmetric, with means close to 0.5 and standard deviations around 0.08. The spread in success rate within one sampling is relatively large -- in other words, the performance of annealing can vary significantly between identical experiments. On the other hand, a comparison of the three samplings reveals no significant difference in the distributions. Therefore, repeating samplings at different times to account for external factors is not necessary for the purposes of our study.

\subsubsection*{The effect of embeddings on annealing performance}
The described experiment was performed using a fixed minor embedding, depicted in Supplementary Figure 1a. It is well established that the choice of embedding can have a significant effect on the performance of annealing. To study this with our approach, we vary the embedding of the problem in two steps. First, we merely translate the original embedding within the lattice of the quantum computer, while keeping its shape intact. We then perform three samplings using the three different, translated embeddings, that are illustrated in Supplementary Figure 1a-c. The corresponding distributions of the success rate are depicted in Figure 1c. Finally, we generate three new minor embeddings that are both translated and have a different topology as compared to the original embedding. They are illustrated in Supplementary Figure 1d-f. The respective distributions of success rate are depicted in Figure 1d.  
We observe significant differences between the success rate distributions of the translated embeddings, and even greater differences between the distributions of completely different embeddings. The fact that embeddings with an identical topology but different positioning in the D-Wave lattice yield significantly different distributions, leads to the conclusion that the topology of the embedding is not the only important factor for performance. In other words, the calibration error is specific to each individual physical qubit. On the other hand, the topology of the embedding has an important effect as well, as evidenced by the larger differences in Figure 1d for embeddings that are both translated and have a different topology. 
There is one final important observation we can make from Figure 1b-d. Because the annealing performance varies significantly between different embeddings, this hints towards a simple but possibly effective method of improving the performance -- generating and trying out different embeddings. However, it must be emphasized that due to the high level of random noise present, as well as the stochastic nature of quantum measurements, experiments that seek to evaluate annealing performance must be repeated many times to get reliable performance estimates.

\begin{figure*}
        \centering
        \includegraphics[width = 0.8\textwidth]{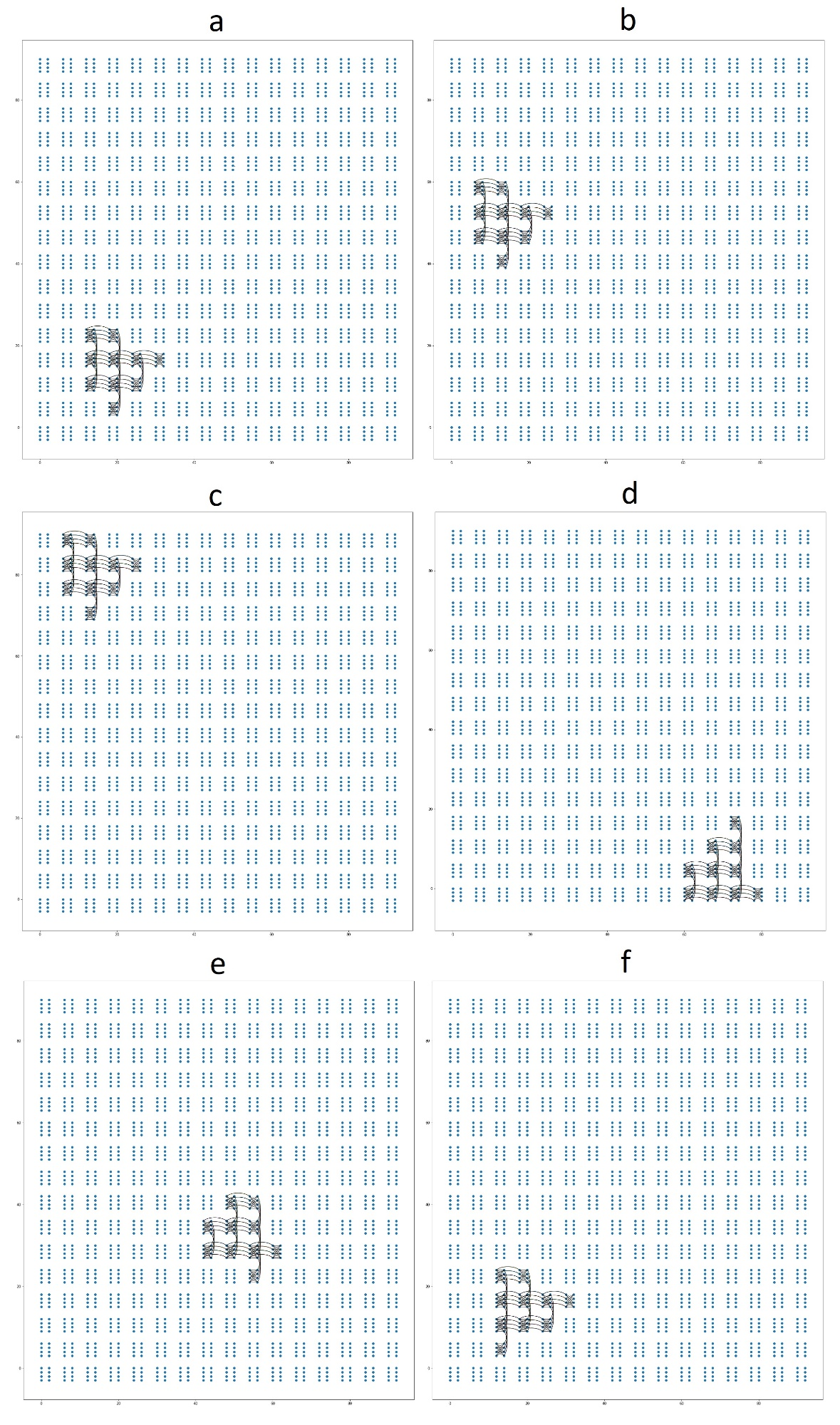}
        \caption{The embeddings used in the experiments in Figure 1b-d. Each blue dot represents a physical qubit in the quantum annealer. Black lines depict the chains of qubits that constitute the embedding.}
        \label{SIfig1}
\end{figure*}

\subsection*{Annealer calibration}
The main problem studied in this work is how to calibrate a quantum annealer to improve annealing performance and counteract systematic noise. As demonstrated in the previous section, simply altering or translating the embedding can remarkably improve performance. This behaviour has already been observed and, indeed, the D-Wave documentation advises trying out different embeddings if performance is not satisfactory. Another method found that not all chains are equally susceptible to noise, due to the specifics of the embedding topology, and that correcting the chain strength parameter for individual chains improves performance. Our experiments, the results of which are depicted in Figure 1b-d, show that this may not be enough and that better improvement may be achieved by calibration at the level of individual qubits. To that effect, we consider two approaches.

\subsubsection*{Anneal offsets}
The first approach involves varying the anneal offset for each physical qubit. This parameter represents a temporal offset in the annealing paths of the qubits, so that some are annealed slightly before others. Anneal offsets could provide the necessary leverage to counteract the systematic noise of each physical qubit and maximize the success rate. Supplementary Figure 2 shows the behavior of annealing performance when changing annealing offsets. The success rate can vary from a relatively high value to almost 0 with large changes to annealing offsets. However, in the best case scenario annealing performance stays the same, and in most cases it worsens. We conclude that annealing offsets are best left at their default, since there is not much room left for improvement. Consequently, this approach is not appropriate for annealer calibration, at least for our sample problem and topology.
This is consistent with the documentation provided by D-Wave, which states that anneal offsets have a large impact on performance only in the case where the chain lengths of different logical qubits vary significantly. In our fully connected graph form of the embedding, the chains have the same lengths. This is therefore a possible avenue of further research, where other kinds of embeddings are considered, with chains of different lengths.

\begin{figure*}
        \centering
        \includegraphics[width = 0.95\textwidth]{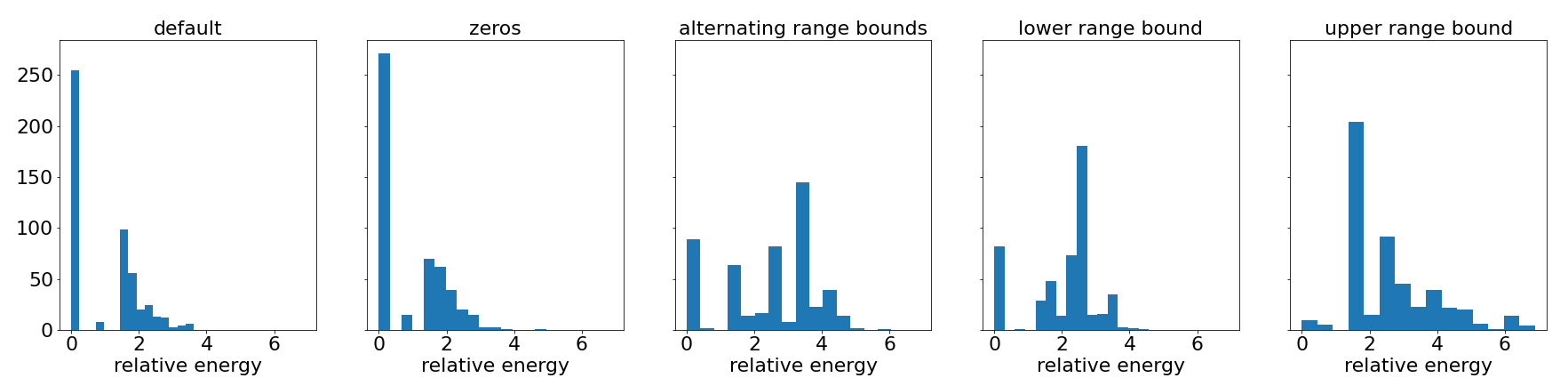}
        \caption{Comparison of results obtained with different annealing offsets. Each histogram represent the results gathered from 500 independent runs on the quantum annealer, all with the same predetermined offsets. From left to right, the annealing offsets were set to "default", the default calibration provided by D-Wave, "zeros", where all offsets were set to zero, "alternating range bounds", where offsets vary from the minimum to the maximum possible value from qubit to qubit, "lower range bound", where all offsets are set to their lowest possible value, and "upper range bound", where all the offsets are set to their highest possible value.}
        \label{SIfig2}
\end{figure*}

\subsubsection*{Hamiltonian corrections}
In the second approach, we consider the QUBO Hamiltonian Q to be the input to the quantum annealer. In the process of translation to the physical system, the Hamiltonian is corrupted by both random and systematic noise. Can we learn to correct the input so as to produce the correct output, and in this way counteract systematic noise? We introduce small corrections to the input Hamiltonian:
$$Q=Q_0+dQ,$$
where Q is the input to the D-Wave sampler, $Q_0$ is the original Hamiltonian that corresponds to the physical problem we are solving, and $dQ$ is the calibration matrix, with $|dQ_{ij}| \ll|Q_{0_{ij}}|$, or alternatively, $||dQ|| \ll ||Q||$, where $||Q||$ indicates the matrix norm of Q.  We vary the elements of $dQ$ and try to maximize the success rate. 
The elements of Q correspond to the coupling strength between pairs of logical qubits. As an upper-triangular matrix, Q has  $D(D+1)/2$ values for a fully connected system of DxD logical qubits. In this work, we study the calibration of the sample problem on three system sizes: 4x4, 5x5 and 6x6,  with the corresponding dimensionalities of 136, 325 and 666.
The calibration of the success rate through varying the matrix elements is a very challenging problem of noisy function optimization in an extremely high-dimensional space. Since we know nothing of the landscape of the systematic noise function, we employ a data-driven approach. We first conduct a large number of experiments by randomly sampling $dQ_{ij}$ values and then employ machine learning methods for regression to analyze the data.

\subsection*{Sampling the $dQ$ space}
Before we can sample the $dQ$ matrices for our dataset, we must properly define the space we are interested in.
When we invoke the D-Wave sampler for a given Q, the D-Wave system performs a few modifications on the matrix. First the matrix is normalized so that the largest absolute value of its elements is 1. Next, the system translates the matrix from the logical qubit representation to the physical qubit representation, using the embedding we provide. Finally, the system constructs qubit chains by assigning the chain strength parameter to the couplings between physical qubits in each chain. 
The normalization of the input after our corrections could compromise our data and confuse modelling. To avoid it, we normalize the $Q_0$ matrix ourselves, and place a constraint on the sampling of $dQ$ so as to ensure that the normalization is not broken by $dQ$. We define the absolutely largest element of $Q_0$ as  
$$ Q_{\text{max}} = \max_{ij}|Q_{0_{ij}}|,$$
and derive the allowed range of each matrix element as
$$\eta \left( -1 - \frac{Q_{ij}}{Q_{\text{max}}} \le dQ_{ij} \right) \le \eta \left( 1 - \frac{Q_{ij}}{Q_{\text{max}}} \le dQ_{ij} \right),$$
where $0 \le \eta \le 1$ denotes the fraction of the range of each matrix element we wish to consider.

The above formulation, with $\eta=1$, describes the entire space of possible calibration matrices $dQ$, constrained only by normalization. We have assumed that the corrections of the input Hamiltonian are small. We control the size of the corrections through the parameter $\eta$. Choosing a very small 
$\eta$ risks cutting the best possible $dQ$ out of our search space. On the other hand, a large $\eta$ vastly increases the size of our search space and consequently the QPU time demands. More importantly, a large $\eta$ carries the risk that we wander so far away from $\eta$ that the problem considered by the quantum computed no longer bears any connection to the physical problem we are studying. For example, the space of all Q also contains a Hamiltonian that represents a degenerate system with all energies equal to the ground state energy of the system we are interested in. Such a Q would achieve seemingly perfect annealing performance, but would be entirely useless in practice. In order to ensure that our calibrated system is still solving the correct problem, it is paramount to keep the corrections of the Hamiltonian relatively small.
For our experiments we used the value $\eta=0.05$ as an educated guess.
Now that we have defined the boundaries of the high-dimensional space we are to explore, we randomly sample it to generate the set of calibration matrices to be evaluated on the quantum annealer. We employ latin hypercube sampling, a method for random sampling from multidimensional distributions that ensures good coverage of the hyperspace, while keeping the required number of samples low. We study the annealing performance of three samplings, one for each of the three system sizes. In each sampling, we evaluate 10000 calibration matrices, each comprising 500 reads from the quantum annealer.

\subsection*{Analysis of the data}
\begin{figure*}
        \centering
        \includegraphics[width = 0.9\textwidth]{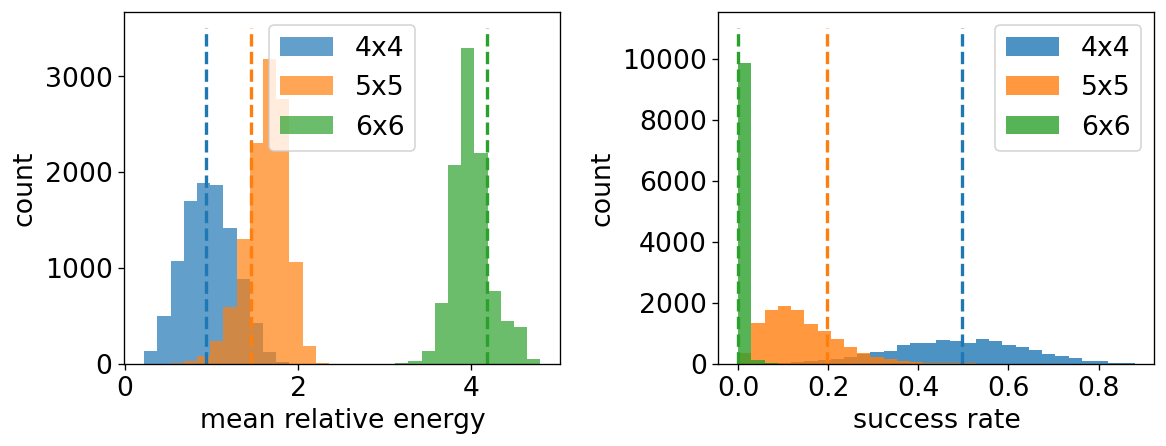}
        \caption{Distributions of the mean relative energy (left) and the success rate (right), obtained by sampling the space of calibration matrices $dQ$, for three system sizes. For each system size, 10000 matrices were sampled. Each sample is composed of 500 reads from the quantum annealer, which are summarized with the mean relative energy and success rate statistics. The mean relative energy and success rate for the original Hamiltonian with no calibration ($Q = Q_0$) are depicted with a dashed line of the corresponding color.}
        \label{SIfig3}
\end{figure*}
The three constructed datasets are summarized in Supplementary Figure 3. There are significant differences in annealing performance for different system sizes. For the 4x4 system, the mean success rate is close to 50\%, for the 5x5 around 14\% and for the largest 6x6 system, the mean success rate is only around 0.5\%. In fact, for many calibration matrices for the 6x6 system, not even a single read ended up in the ground state. A comparison between the distributions of the success rate and mean relative energy reveals that while the success rate is a nicer and more intuitive metric, its usefulness drops sharply when the annealing performance is very poor. In such cases, it is better to use the mean relative energy as the measure of annealing performance.
In Supplementary Figure 3, the annealing performance for the uncalibrated Hamiltonian ($Q = Q_0$) is depicted by a dashed line for each system. We can see that, for all system sizes, a good fraction of the evaluated calibration matrices achieves better annealing performance than the original matrix. This observation is surprising, since we expected it to be very difficult to find a calibration that improves performance in the high dimensional $dQ$ spaces.
In fact, taking the calibration matrix that yields the best annealing performance for each system size, we arrive at the results summarized in Table 1 in the main text.
The results are already impressive and demonstrate the effectiveness of error correction by corrections to the input Hamiltonian. Based on these results, we define the first, simple but effective, error correction algorithm – the argmax strategy, described in Algorithm 1 in the main text. 

\subsection*{Performance curve estimation through bootstraped subsampling}
When analyzing the efficiency of our error correction algorithms, it is very informative to obtain an estimation of how the method performs at various sample sizes. In Figure 2a, we depict an estimation of the success rate achieved by the argmax strategy for a given size of the $dQ$ sampling. 
In typical optimization settings, we might plot the performance curve by simply taking the best measured result at a given step in the optimization algorithm. In our case, however, we are interpreting the problem as independent random sampling.  We can thus make a better estimate of the performance curve through a process of repeated bootrapped subsampling.
For a given sample size $1 \le N_i \le N; N = 10000$, we randomly choose $N_i$ calibration matrices out of all evaluated calibration matrices, and apply the argmax strategy to the obtained subsample. We repeat this procedure 1000 times and compute the mean and standard deviation of the resulting values of the success rate. By varying $N_i$ and performing this procedure for each system size, we obtain the curves depicted in Figure 2a in the main text.

\subsection*{Machine learning method evaluation}
We evaluate the predictive performance of models obtained by different machine learning methods through 5-fold cross validation. We randomly partition the dataset into 5 fractions. Models are trained on four of the fractions (training set) and attempt to predict the fifth (testing set) fraction. In  each of the five iterations, a different fraction is used as the testing set. In the end, the predictions are pooled together and the prediction performance is evaluated. As the metric of comparison, we use the coefficient of determination, defined as
$$R^2 = 1 - \frac{\sum_i (\Tilde{y}_i - \Bar{y})^2} {\sum_i (y_i - \Bar{y})^2}),$$
where $\Tilde{y}_i$ indicates the predicted annealing performance of the i-th calibration matrix, $\Bar{y}$ is the mean annealing performance for the testing set and $y_i$ denotes the measured annealing performance of the i-th calibration matrix.
The coefficient of determination can be interpreted as the portion of variance in the data explained by the model. A perfect model would achieve $R^2  = 1$, whereas $R^2  < 0$ indicates the model is worse than simply taking the average over the test set.
In the analysis of the data, we observed high (anti) correlation between mean energy and success rate as measures of annealing performance. The success rate is problematic for the 6x6 system, since many calibration matrices do not end up with any reads in the ground state. For this reason, we have chosen mean relative energy as the measure for annealing performance when training predictive models, even though the success rate offers easier interpretation.

\subsection*{Machine learning method details}
The ``no free lunch" theorem in machine learning states that no single learning algorithm can simultaneously be the best for every problem. We have tried out and evaluated a number of different algorithms for learning models that take the entries of a calibration matrix as the input and predict the mean relative energy of a sample. Below we give details on the (hyper)parameter settings of the employed models. For details on the machine learning algorithms, we refer the reader to the work, cited in the main text. We used the Python Scikit-learn\textsuperscript{1} implementations for all machine learning methods except for neural networks, for which we used Tensorflow\textsuperscript{2}.

\footnote{\textsuperscript{1}Pedregosa, Fabian, et al. "Scikit-learn: Machine learning in Python." \textit{The Journal of machine Learning research} 12 (2011): 2825-2830.}
\footnote{\textsuperscript{2}Abadi, Martín, et al. "Tensorflow: A system for large-scale machine learning." \textit{12th {USENIX} symposium on operating systems design and implementation ({OSDI}} 16). 2016.}

\subsubsection*{Linear regression with L1 regularization - Lasso}
\vspace*{-\baselineskip}
\begin{table}[H]
   \centering
   \begin{tabular}{lrrr}\\
        \toprule
       	  \textbf{\textbf{Parameter/system size}} & \text{4x4} & \text{5x5} & \text{6x6} \\
       	  \midrule
       	  \textbf{Regularization $\alpha$} & 1e-5 & 7e-6 & 2e-5 \\
       	\bottomrule
   \end{tabular} 
\end{table}

\subsubsection*{Linear regression with L2 regularization – Ridge}
\vspace*{-\baselineskip}
\begin{table}[H]
   \centering
   \begin{tabular}{lrrr}\\
        \toprule
       	  \textbf{\textbf{Parameter/system size}} & \text{4x4} & \text{5x5} & \text{6x6} \\
       	  \midrule
       	  \textbf{Regularization $\alpha$} & 0.16 & 0.53 & 1.0 \\
       	\bottomrule
   \end{tabular} 
\end{table}

\subsubsection*{Support vector machines}
\vspace*{-\baselineskip}
\begin{table}[H]
   \centering
   \begin{tabular}{lrrr}\\
        \toprule
       	  \textbf{\textbf{Parameter/system size}} & \text{4x4} & \text{5x5} & \text{6x6} \\
       	  \midrule
       	  \textbf{Kernel type} & RBF & RBF & RBF \\
       	  \textbf{Regularization $C$} & 1.0 & 2.9 & 5.9\\
       	\bottomrule
   \end{tabular} 
\end{table}
RBF stands for radial basis functions. Other parameters were left at their default values.

\subsubsection*{Random forests}
All parameters, such as the number of trees in the forest, were left at their default values.

\subsubsection*{Neural networks}
For all system sizes, we used a network with 3 fully-connected hidden layers with the ReLu activation function and an output layer with linear activation. We used the Adam optimizer, MSE loss and 50 epochs of training. 
We also considered regularizing the network weights, but in our experiments this only adversely affected predictive performance.

\begin{table}[H]
   \centering
   \begin{tabular}{lrrr}\\
        \toprule
       	  \textbf{\textbf{Parameter/system size}} & \text{4x4} & \text{5x5} & \text{6x6} \\
       	  \midrule
       	  \textbf{First layer size} & 100 & 300 & 600\\
       	  \textbf{Second layer size} & 50 & 100 & 100\\
       	  \textbf{Third layer size} & 20 & 20 & 20\\
       	  \textbf{Learning rate} & 5e-5 & 9e-5 & 1e-3\\
       	\bottomrule
   \end{tabular} 
\end{table}

\subsubsection*{Other machine learning methods}
We have considered other machine learning approaches, such as k-nearest-neighbors, linear regression with a quadratic basis, logit regression, and extreme gradient boosting. These approaches either achieved performance significantly worse than the ones reported, or were too computationally demanding, so we dropped them from consideration.

\subsection*{Predictive strategy details}
Since linear regression is the simplest and most robust of the models we have tried and it still achieved competitive predictive performance, we have chosen it for implementing the predictive strategy. We now have a model that predicts mean relative energy for a given calibration matrix $dQ$ and we can use it as a stand-in for experiments on the quantum annealer. To find the best calibration, we use a method for numerical minimization (in our case, differential evolution as implemented in Python-Scipy\textsuperscript{3} ) to find the minimum of the mean relative energy. 
Minimization requires bounds on the high-dimensional space. Since we know only that the calibration matrix should be small in comparison to $Q_0$, we once again invoke the $dQ_{ij}$ ranges defined earlier and vary the range fraction $\eta$. For each value of $\eta$, we minimize the model to obtain a candidate calibration matrix. 
We then evaluate the matrix on the quantum annealer, repeating the measurement 10 times, each time with 500 reads, and compute the mean and standard deviation of the measured success rate. By performing this procedure for each value of $\eta$ (using the same model) and for each system size (using the respective models), we obtain the curves depicted in Figure 2b-d.

\footnote{\textsuperscript{3}Virtanen, Pauli, et al. "SciPy 1.0: fundamental algorithms for scientific computing in Python." \textit{Nature methods} 17.3 (2020): 261-272.}

\subsection*{Learning curve estimation}
In Figure 1a, we estimate the dependence of annealing performance improvement of the argmax strategy on the size of the dataset. Obtaining an analogous curve for the predictive strategy would be costly in terms of QPU time, since each $dQ$ candidate needs to be evaluated on the quantum annealer, in contrast to the argmax strategy, where we are using measurements directly. However, we can use the existing data to estimate the dependence of the predictive performance $R^2$ of the model on the size of the dataset. Since a model that predicts well helps us find better calibration matrices, we can expect the improvements in annealing performance to be strongly positively correlated with predictive performance.
We follow the same repreated bootrap subsampling procedure described earlier, but this time we train predictive models on the subsampled data. After repeating this 100 times for each sample size $N_i$, we compute the mean and standard deviation of $R^2$ at each $N_i$. The resulting learning curve for linear regression is depicted in Supplementary Figure 4.

Based on the analysis, we can expect the predictive strategy to perform well for the 4x4 system with a sample size as small as 1000-2000 $dQ$ matrices, with the coefficient of determination saturating for larger sample sizes. For the 5x5 system, between 3000 and 5000 samples would be the most cost-effective. For the 6x6 system, even 10000 samples are not enough for the predictive strategy to perform well.

\begin{figure}
        \centering
        \includegraphics[width = 0.5\textwidth]{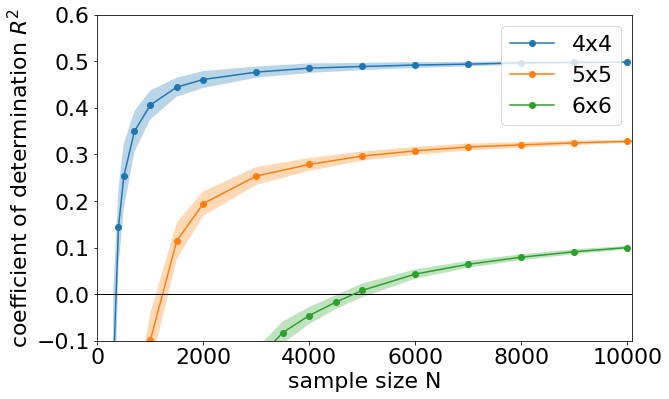}
        \caption{The estimated learning curve for linear regression, obtained by repeated bootrap subsampling.}
        \label{SIfig4}
\end{figure}

\end{document}